\documentstyle[prd,aps,epsfig]{revtex}
\begin{document}
\title{Possible test of local Lorentz invariance from $\tau$ decays}
\author{L. Anchordoqui\and M. T. Dova \and D. G\'omez Dumm \and P. Lacentre}
\address{Laboratorio de F\'{\i}sica Te\'orica, Departamento de F\'{\i}sica,
\\ U.\ N.\ L.\ P., c.\ c.\ 67, 1900 La Plata, Argentina.}

\maketitle

\begin{abstract}
We analyze the possibility of testing local Lorentz invariance from
the observation of tau decays. Future prospects of probing distances
below the electroweak characteristic scale are discussed.
\end{abstract}

\begin{center}
{\em Journal Reference: Zeitschrift f\"ur Physik C {\bf 73} (1997) 465.}
\end{center}


{\em {\bf Introduction.}} 
The question of the existence of a universal length, below which our 
present notion of flat spacetime geometry is not valid, is of great
theoretical importance \cite{will}. In this respect, it is important
to point out that the current success of the Standard Model (SM) in
describing the electroweak processes serves as a constraint for any
observable breakdown of local Lorentz invariance (LLI), which should
appear as a small deviation from the SM predictions. In this way, any
search for LLI nonconservation effects will require to test distances
below the electroweak scale, i.e. of order $10^{-16}$ cm or less.

In order to get some insights into this subject, one possible way is
to consider the experimental information on the decays of particles
which move at high velocities in the lab system. The comparison
between the measured lifetimes at these energies and the
corresponding values in the rest frame would provide a direct test of
the time dilation formula. In this direction, a proposal has been
recently presented by Freeman {\em et al.} \cite{HOJ} to measure the
lifetimes of 530 GeV charged pions at Fermilab.

In this article, we concentrate on the possibilities of testing LLI
from the observation of tau lepton decays. In particular, we consider
the strongly relativistic tau pairs produced at LEP-I, taking profit
of the increasing statistic and excellent experimental precision of
the tau lifetime measurement. By comparing this lifetime value with
the low energy results, it is possible to determine an upper bound
for the LLI-breaking characteristic length. In addition, we consider
the possibility of testing LLI from the measurement of particular
$\tau$ decay branching ratios, instead of dealing with the total
decay width. The future prospects of probing distances below the
electroweak characteristic scale are also discussed.

\hfill

{\em {\bf Framework.}} 
The idea of testing LLI from particle decay rates is
not new. Indeed, this possibility was suggested many years ago by
R\'edei \cite{red}, who studied possible LLI breakdown effects on the
decay of charged pions and muons. Now, considering the extraordinary
evolution of the experimental knowledge, it is worth to recover this
proposal including present data on particle lifetimes. We will use
this framework extended to the $\tau$ lepton decays.

Following the scheme of R\'edei, we begin by introducing into the
theory a timelike unit vector $n$ which characterizes the chosen
reference frame \cite{bolo}. As a consequence, the transition
amplitudes become (in general) not invariant under a Lorentz
transformation of the particle states if the reference system is kept
unchanged. The theory does not exhibit Lorentz invariance in the
so-called ``active'' sense.  However, one still has Lorentz
invariance in the ``passive'' sense, i.e., under a simultaneous
transformation of the particle variables and the vector $n$ (this is
necessary in order to guarantee that the probabilities remain
independent of the velocity of the observer).

We will study the main tau decay partial widths considering in each
case a noncausal effective Hamiltonian of the type proposed by
R\'edei,
\begin{equation}
\int {\cal H}_{\rm int}\, d^4x =  
\frac{G_F}{\sqrt{2}} \int\!\!\int d^4x\, d^4y \,\,\bar{\nu}_{\tau}(x)
\gamma^{\mu} (1-\gamma_5) \tau(x) 
\,F(x-y)\, J_{\mu}(y)\;,
\label{H}
\end{equation}
where the current $J_{\mu}$ depends on the decay channel, and the
noncausality is introduced through the form factor $F(x)$. The latter
is given by
\begin{displaymath}
F(x) = \frac{3}{4 \pi \alpha^3} \, \delta(n\cdot x) \, \Theta
(\alpha^2 + x^2 - (n\cdot x)^2 \,)
\end{displaymath}
and satisfies $F(x) \rightarrow \delta(x)$ in the limit $\alpha
\rightarrow 0$. In the lab system (which can be considered at rest
with respect to the surrounding macroscopic bodies\footnote{Most of
the low velocity experiments which test LLI consider
a preferential reference frame comoving at 350 km/s with the
center of mass of the Universe (as seen from the dipole asymmetry of
the 3 K background radiation \cite{fc}). However, this cosmic motion
can be neglected for our purposes in view of the high LEP-I
velocities.}), we may take \mbox{$n = (\,1\,,\,\vec{0}\,)$}. Then the
form factor reduces to
\begin{equation}
F(x) = \frac{3}{4 \pi \alpha^3} \, \delta(x_0) \, \Theta (\alpha^2 
- |\vec{x}|^2)\;,
\end{equation}
where it becomes clear that $\alpha$ represents a characteristic
length below which the noncausality effects will take place.

We are interested in computing the decay amplitudes in the lab system
for strongly relativistic tau leptons. The $\tau$ polarization will
be also included, since it can in principle contribute non-negligibly
to the LLI breakdown effects we are looking for (in the case of
LEP-I, the taus are produced with a polarization of
approximately 14\%)\footnote{The spatial component of $n$ is not
zero in the rest frame of the decaying tau.  Therefore, $n$
introduces a spatial anisotropy and the decay probability is in
general a function of the tau polarization direction.}.

Let us first consider the leptonic decays $\tau^- \rightarrow l \,
\bar{\nu}_l \, \nu_{\tau}$ ($l=\mu^-,e^-$) for a $\tau$ lepton with
polarization $P_\tau$. The corresponding decay amplitude is found to
be
\begin{equation}
|{\cal M}|^2 = 64 \; G_F\,\!^2 \; (q_\tau\cdot q_{\nu_l} - m_\tau \;
s\cdot q_{\nu_l}) 
\, (q_\nu\cdot q_l)\;  [\,g(n, q_\tau\! -\! q_{\nu_\tau})\,]^2\;.
\end{equation}
where $s$ stands for the tau polarization 4-vector ($q_{\tau}\cdot
s=0,\, s^2 = -{P_\tau\,\!}^2$) and we have neglected the lepton mass
$m_l$. The function $g(n,Q)$ is the Fourier transform of $F(x)$,
\begin{equation}
g(n,Q) = 1 - \frac{1}{10} \, \alpha^2 \left[(Q\cdot n)^2 - Q^2
\right] + {\cal O}(\alpha^4)\;.
\end{equation}

In order to calculate the partial widths $\Gamma_l(E_{\tau})$, it is
convenient to perform a transformation to the tau rest frame.
However, notice that one has to boost both the particle variables and
the vector $n$ to get the right result. The Fourier transform of
$F(x)$ in the new reference system reads
\begin{equation}
g(n,Q)^{(rest)} = 1 - \frac{1}{10}\,\alpha^2\, \left[
\gamma^2 v^2\, (E_{\nu_{\tau}}-m_\tau)^2
+ E_{\nu_{\tau}}^2 
+ 2\, \gamma^2 v \, (E_{\nu_{\tau}}-m_\tau) \,
E_{\nu_{\tau}} \cos \theta + \gamma^2 v^2 E_{\nu_{\tau}}^2 \cos^2
\theta \right]\;,
\end{equation}
where $\theta$ is the angle between the boost direction and the
3-momentum $\vec q_{\nu_\tau}$ and we have used $\gamma^2 =
(1-v^2)^{-1}$, being $v$ the tau velocity in the lab system. After
integration over the phase space, we find for the leptonic partial
widths
\begin{equation}
\Gamma_l(E_\tau)=\frac{G_F^2\, m_\tau^5}{192\,\pi^3}\,
\gamma^{-1} [1-\alpha^2\delta_l(E_\tau)]\;, \hspace{1cm} l=e,\mu\;,
\label{gamma}
\end{equation}
with
\begin{equation}
\delta_l(E_\tau)=\frac{1}{5}\, E_\tau^2
\left(\frac{2}{15}+ \frac{31}{90}\, v^2+\frac{1}{18}\,P_\tau
v\right)\;.
\label{dlep}
\end{equation}
In the limit $\alpha\rightarrow 0$, we get the relation
\begin{equation}
\Gamma_l(E_\tau)=\gamma^{-1}\Gamma_l(m_\tau)\;,
\end{equation}
i.e. we recover the usual time dilation formula.

We consider now the decays $\tau^-\rightarrow h\nu_\tau$, with
$h=\pi^-,\rho^-,a_1^-$. The contribution of these processes, together
with the leptonic channels treated above, add up to approx.\ 90\% of
the total tau decay width. In analogy with (\ref{gamma}), the
corresponding partial widths for $\tau$ leptons with energy $E_\tau$
can be written as
\begin{equation}
\Gamma_h(E_\tau)=\Gamma_h(m_\tau)\gamma^{-1} [1-\alpha^2
\delta_h(E_\tau)]\;,
\hspace{1cm} h=\pi^-,\rho^-,a_1^-\;,
\end{equation}
where we have assumed $\delta_h(m_\tau)\ll\delta_h(E_\tau)$.
This is justified when we deal with ultrarelativistic taus.
The deviation factors $\delta_h$ can be easily calculated, resulting
\begin{equation}
\delta_h(E_\tau)=\frac{1}{20}\, E_\tau^2 \left[(1-x_h)^2+
\frac{1}{3}\,(1+10\,x_h+x_h^2)\,v^2
+\frac{2}{3}\,P_\tau\lambda_h\,(1-x_h^2)\,v\right]\;,
\label{dhad}
\end{equation}
where $x_h\equiv m_h^2/m_\tau^2$, and $\lambda_h$ is given by
\begin{equation}
\lambda_h\equiv \left\{ \begin{array}{cll}
1 & \mbox{for} & h=\pi^- \\
\frac{1-2\,x_h}{1+2\,x_h} & \mbox{for} & h=\rho^-,a_1^-\;.
\end{array} \right.
\end{equation}
In the same way, the corrected expression for the total $\tau$ decay
rate will read
\begin{equation}
\Gamma(E_\tau) = \sum_i \Gamma_i(E_\tau) = \gamma^{-1}\Gamma(m_\tau)
\left(1-\alpha^2\Delta(E_\tau)\right)
\label{decrate}
\end{equation}
with
\begin{equation}
\Delta(E_\tau)\equiv\sum_i
\frac{\Gamma_i}{\Gamma}\,\delta_i(E_\tau)\;,
\label{delta}
\end{equation}
where the sum extends in principle to all $\tau$ decay channels.
We will neglect here the deviations corresponding to the less
significant branching ratios, restricting ourselves to the
$\delta_i$ contributions given by (\ref{dlep}) and (\ref{dhad}).
Notice that the energy dependence of $\Gamma_i$ and $\Gamma$ in
(\ref{delta}) has been omitted, since it would imply a correction of
order $\alpha^4$ to the total decay rate (\ref{decrate}).

It is important to remark that, within the scheme under
consideration, the branching ratios are no longer fixed numbers. They
depend in general on the energy of the decaying particle. In the case
of the $\tau$ leptons, the above equations conduce to the (order
$\alpha^2$) relations
\begin{equation}
\frac{\Gamma_i(E_\tau)}{\Gamma{(E_\tau)}}
=\frac{\Gamma_i(m_\tau)}{\Gamma{(m_\tau)}}\,
\left[1+\alpha^2(\Delta(E_\tau)-\delta_i(E_\tau))\right]\;.
\end{equation}
In this way, the sole measurement of branching ratios provides
another possible test of LLI.

\hfill

{\em {\bf Numerical analysis.}} As stated above, the experimental data
on the $\tau$ lepton lifetime at both high and low energies provide
an upper bound for the characterstic LLI breakdown length
$\alpha$. In addition, it is possible to estimate how much energy
would be necessary to probe distances below the electroweak scale for
a given accuracy in the $\tau$ lifetime measurements.

Let us first consider the $\tau$ leptons which are produced at
approximately 45 GeV in LEP-I and SLD. From the corresponding experimental
data, the weighted average value of the $\tau$ lifetime is found to be
\cite{altas}
\begin{equation}
\tau_\tau = 291.4\pm 1.6\,\mbox{ fs}\; .
\label{vidalep}
\end{equation}
For low energy $\tau$'s, the lifetime value is less accurate.
The present measurements include a 2.5\% error, which is indeed the
dominant one when determining the bound for $\alpha$. By combination of
the CLEO \cite{cleo} and ARGUS \cite{argus} results, we obtain
\begin{equation}
\tau_\tau = 292\pm 7 \,\mbox{ fs}\; .
\label{vidabaj}
\end{equation}
As can be seen, the data in (\ref{vidalep}) and (\ref{vidabaj}) are
compatible with a null value of $\alpha$. In order to obtain an upper
bound, we consider the modified time dilation formula in eq.\
(\ref{decrate}), with the $\delta_i$ corrections given by (\ref{dlep}) and
(\ref{dhad}). With a 95\% confidence level, we find
\begin{equation}
\alpha\leq 3.3\times 10^{-16} \mbox{ cm}\;,
\end{equation}
where the numerical evaluation of the total deviation factor $\Delta$
has been performed taking the experimental values for the particle
masses and $\tau$ decay branching ratios from Ref.\ \cite{tabla}.

From the expression (\ref{decrate}), one can also examine the future
possibilities of probing even smaller distances. In figure 1, we plot
the value of $\alpha$ that would be sensitive to this test for a
given energy of the decaying tau. The accuracy in the tau lifetime
measurement (this means, for both high and low energies) is included as
a parameter. If we take into account $\tau$-leptons of $\sim 50$ GeV,
it can be seen that this accuracy should be increased up to a 0.5\% in
order to reach a sensitivity to $\alpha$ below the electroweak
characteristic length of $10^{-16}$ cm. For energies of $\sim 100$ GeV,
however, distances beyond this limit could be probed with a
$\tau$ lifetime precision of about 2\%.

Another possible test of the presence of LLI violation can be
obtained by comparing the $\tau$ and $\mu$ leptonic partial widths
\cite{lep}. Let us consider the ratio
\begin{equation}
R(E_\tau)\equiv\frac{\Gamma(\tau\rightarrow\nu_\tau e \bar\nu_e)}{
\Gamma(\mu\rightarrow\nu_\mu e \bar\nu_e)}
=B_e(E_\tau)\,\frac{\tau_\mu(E_\mu)}{\tau_\tau(E_\tau)}\; ,
\label{rdef}
\end{equation}
where $B_e$ is the branching ratio for the $\tau\rightarrow
\nu_\tau e\bar\nu_e$ decay and $E_\tau$ is the energy of the decaying
$\tau$ in the lab system. Within our framework, it is easy to show that
\begin{equation}
R(E_\tau)=\left(\frac{g_\tau}{g_\mu}\right)^2 \!\left(\frac{m_\tau}{
m_\mu}\right)^5 \!(\Delta_W \Delta_\gamma)^{-1} (1-\alpha^2
\delta_l(E_\tau))\; ,
\label{rteo}
\end{equation}
where $g_l$ stands for the coupling constant corresponding to the weak
charged current of a lepton $l$. The factors $\Delta_\gamma\simeq
1+8.6\times 10^{-5}$ and $\Delta_W\simeq 1-3.0\times 10^{-4}$ come from
the inclusion of radiative corrections to the tree level decay amplitudes.
We have not included the correction $\delta_l(E_\mu)$, which
has been assumed to be much smaller than $\delta_l(E_\tau)$. From
(\ref{rdef}) and (\ref{rteo}), we finally have
\begin{equation}
\left(\frac{g_\tau}{g_\mu}\right)^2 (1-\alpha^2\delta_l(E_\tau)) =
\left(\frac{m_\mu}{m_\tau}\right)^5 \frac{\tau_\mu}{\gamma^{-1}
\tau_\tau(E_\tau)}\,B_e(E_\tau)\,\Delta_W \Delta_\gamma\; ,
\label{univ}
\end{equation}
i.e. the presence of the LLI-breakdown parameter $\alpha$ would be
observed as a violation of the $\tau-\mu$ universality.

In reference \cite{lep}, the right hand side of (\ref{univ}) has been
evaluated using the experimental values of the masses and decay rates.
Setting $\alpha=0$, the LEP measurements for $\tau$ decays conduce to
\begin{equation}
\frac{g_\tau}{g_\mu} = 0.9954\pm 0.0043\; .
\label{ratio}
\end{equation}
Within the LLI-violating scheme we are considering, the accurate value in
the above expression, together with the requirement of universality
$g_\tau=g_\mu$, provide a new upper bound for
the parameter $\alpha$. In figure 2, we plot the value of the ratio
$g_\tau/g_\mu$ that is obtained from eq.\ (\ref{univ}) for values of
$\alpha$ from $10^{-17}$ to $10^{-15}$ cm. The solid curve
corresponds to the average experimental values of the parameters entering
the right hand side, while the dashed lines take into account a 2$\sigma$
deviation. Within this confidence level, the agreement with universality
conduces to
\begin{equation}
\alpha\leq 2.3\times 10^{-16} \mbox{ cm}\; .
\end{equation}

Let us finally stress the fact that measurements of branching ratios at
high and low energies can provide accurate tests of LLI. In this
sense, the $\tau$ leptons represent an ideal laboratory, in view of the
various significant decay channels which can be observed. The prospects
are also encouraged by the present improvement in these measurements at
LEP \cite{aleph} and the proposal of building a Tau-Charm factory
\cite{chinos}, where the expected precision on the values of branching
ratios will be far below 1\% in one-year data sample.

\acknowledgements 
We would like to thank L.\ Epele, H.\ Fanchiotti,
 C.\ Garc\'{\i}a Canal and J.\ Swain for useful discussions and comments.
This work was partially supported by CONICET (Argentina).

\clearpage

\centerline{\large\bf Figures}

\hfill

\begin{figure}[htbp]
\begin{center}
\epsfig{file=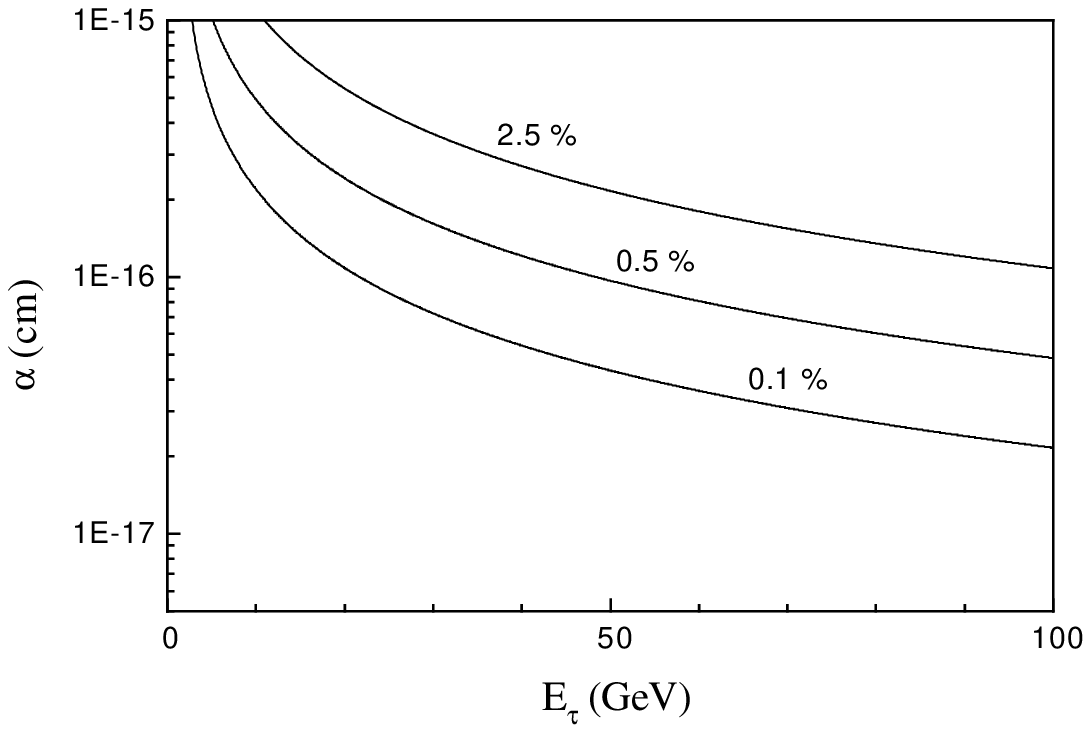}
\end{center}
\vspace{1cm}
\caption{Distance that is probed by the LLI-test, as a function of
the $\tau$ lepton energy. Different accuracies in the $\tau$
lifetime measurements are considered.}
\end{figure}

\hfill

\begin{figure}[htbp]
\begin{center}
\epsfig{file=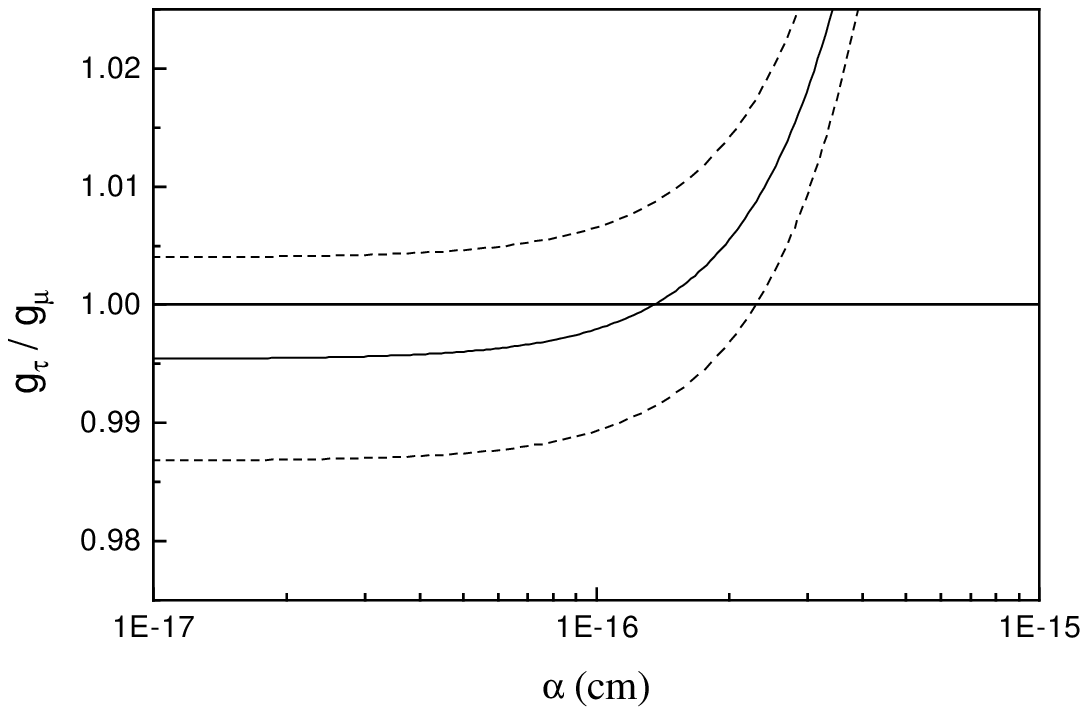}
\end{center}
\vspace{1cm}
\caption{Ratio $g_\tau/g_\mu$ calculated from measured values of $\tau$
and $\mu$ masses and leptonic decays. Solid and dashed lines represent
average and 2$\sigma$ values respectively.}
\end{figure}


\begin{references}

\bibitem{will} M.\ P.\ Haugan, Ann.\ Phys.\ (NY) {\bf 118} (1979)
156.

C.\ M.\ Will, Phys.\ Rep.\ {\bf 113} (1984) 345.

\bibitem{HOJ} {\em P874: Proposal to Measure the $\pi^+$ and
${\pi^-}$ Lifetimes at High Energy}, J. Freeman, S. Geer, C. Hojvat,
C. Johnstone, J. Streets, D. Striley, L. DeMortier, N. D. Giokaris,
D. M. Khazins, S. Oh, T. Phillips, C. Wang, G. Snow and C. Lundstedt,
private communication.

\bibitem{red} L.\ B.\ R\'edei, Phys.\ Rev.\ {\bf 145} (1966) 999;
{\bf 162} (1967) 1299.

\bibitem{bolo} D.\ I.\ Blokhintsev, Phys.\ Lett.\ {\bf 12} (1964)
272.

\bibitem{fc} G.\ F.\ Smoot, M.\ V.\ Gorenstein, Phys.\ Rev.\ Lett.\
{\bf 39} (1977) 898.

\bibitem{altas} OPAL Collaboration, R.\ Akers {\em et al.}, Phys.\
Lett.\ B {\bf 338} (1994) 497.

ALEPH Collaboration, D.\ Buskulic {\em et al.}, CERN preprint
CERN-PPE/95-128 (1995).

DELPHI Collaboration, P.\ Abreu {\em et al.}, CERN preprint
CERN-PPE/95-154 (1995).

M.\ Biasini (L3 Collaboration), Nucl.\ Phys.\ B (Proc.\ Suppl.) {\bf 40}
(1995) 331.

SLD Collaboration, K.\ Abe {\em et al.}, Phys.\
Rev.\ D {\bf 52} (1995) 4828.

\bibitem{cleo} C.\ White, Nucl.\ Phys.\ B (Proc.\ Suppl.) {\bf 40}
(1995) 311.

\bibitem{argus} ARGUS Collaboration, H.\ Albretch {\em et al.}, DESY
preprint DESY 96-015 (1996).

\bibitem{tabla} Particle Data Group, Phys.\ Rev.\ D {\bf 54} (1996)
1.\

\bibitem{lep} M.\ Davier, Nucl.\ Phys.\ B (Proc.\ Suppl.) {\bf 40}
(1995) 395.

\bibitem{aleph} ALEPH Collaboration, D.\ Buskulic {\em et al.},
CERN preprint CERN-PPE/95-140 (1995); {\em ibid.} CERN preprint
CERN-PPE/95-127 (1995).

\bibitem{chinos} J.\ Kirkby, CERN preprint CERN-PPE/94-037 (1994).

A.\ Pich, Valencia preprint FTUV/95-43, IFIC/95-45 (1995).

\end{references}
\end{document}